\documentclass[12pt]{article}%

\usepackage{graphicx}
\usepackage[utf8]{inputenc}
\usepackage{color}
\usepackage{hyperref}
\usepackage{epsfig}
\usepackage[affil-it]{authblk}

\graphicspath{{./Fig/}}

\begin{document}

\title{Cascading collapse of online social networks}


\author[1,2,*]{János Török}
\author[1,2]{János Kertész}

\affil[1]{Center for Network Science, Central European 
University, N\'ador u. 9, H-1051 Budapest, Hungary}
\affil[2]{Department of Theoretical Physics, Budapest
University of Technology and Economics, H-1111 Budapest, Hungary}

\affil[*]{torok@phy.bme.hu}

\maketitle

\begin{abstract}
Online social networks have increasing influence on our society, they
may play decisive roles in politics and can be crucial for the fate of
companies. Such services compete with each other and some may even
break down rapidly. Using social network datasets we show the main
factors leading to such a dramatic collapse. At early stage mostly the
loosely bound users disappear, later collective effects play the main
role leading to cascading failures. We present a theory based on a
generalised threshold model to explain the findings and show how the
collapse time can be estimated in advance using the dynamics of the
churning users. Our results shed light to possible mechanisms of
instabilities in other competing social processes.
\end{abstract}

Growth as a main route to the formation of complex networks have been
studied in great detail
\cite{Barabasi_1999,Kleinberg_1999,Leskovec_2005,Newman_2010}. Much
less effort has been devoted to the understanding of the complex
process of the decline of networks  \cite{Saavedra_2008}. In this
context the effect of random failures and intentional attacks
\cite{Albert_2000, Cohen_2000, Cohen_2001, Morone_2015} has been
investigated and recently the enhanced vulnerability and cascading
breakdown of interdependent networks \cite{Buldyrev_2010} or k-core
percolation \cite{Goltsev_2006} were shown.  The disintegration of
real networks is usually a consequence of an interplay between
endogenous and exogenous factors and its understanding is of major
interest for a series of important questions like the decay of living
organisms, the disintegration of social networks or the loss of market
share in economic competition.

Social contagion \cite{Rogers_2003,Pastor-Satorras_2015} like adoption
of opinions, behavioural patterns, emotions
\cite{suvakov2012online,tadic2017agent,mitrovic2011quantitative,tadic2013co}
or innovations can be considered as growth of a network of adopters on
the top of an underlying network namely that of social interactions.
Under some circumstances this process is surprisingly rapid. Social
pressure plays a pivotal role in this context: People are influenced
in their decisions by the opinions of their
peers\cite{suvakov2012online,tadic2017agent}. This effect is
captured in the so called threshold models \cite{Granovetter_1978,
Watts_2002,Easley_2010}, which assume that a person becomes adopter,
when the ratio of her already adopting, related peers have reached a
critical level characteristic to her sensitivity. 

The above spreading processes are constructive in the sense that as a
result a network of adopters emerges. The study of such processes have
been boosted by the abundance of information communication data
\cite{Centola_2010,Lerman_2010,Gonzales-Baylon_2011,Weng_2013,
Karsai_2014,Karsai_2016,andjelkovic2015hierarchical} and has lead to a
deeper understanding of how adopter networks emerge due to complex
social contagion
\cite{Watts_2002,Gleeson_2007,Centola_2010,Singh_2013,Gleeson_2011,Gleeson_2013,Karsai_2014,Ruan_2015,tadic2013co}.
However, there can be an opposite process, when people forming a
network of users of some technology or service leave this network. The
reason can be getting uninterested in the service or churning to
another provider \cite{Kim_2004}. Are there collective effects in this
process as well, with the possibility of a dramatic collapse of the
entire adopter network? The first aim of our paper is to use the data
to show that this can indeed be the case and to present a model, which
is able to describe the time dependence of the decomposition of such a
networks. The second aim of this paper is to verify the model if it is
able to predict in advance the collapse of a site.

After the success of Facebook (\url{https://facebook.com}) a
particularly keen competition has developed among Online Social
Networks (OSN-s) resulting in the defeat of a number of earlier
popular sites. We will focus in this study data on {\em
iWiW}\cite{iwiw}, which used to be the most successful Hungarian OSN.
Results will also be shown for a much smaller OSN the Gowalla
available from Large Stanford Dataset Collection~\cite{snapnets}.

\section*{Empirical results}

\begin{figure}
\includegraphics[width=.95\textwidth]{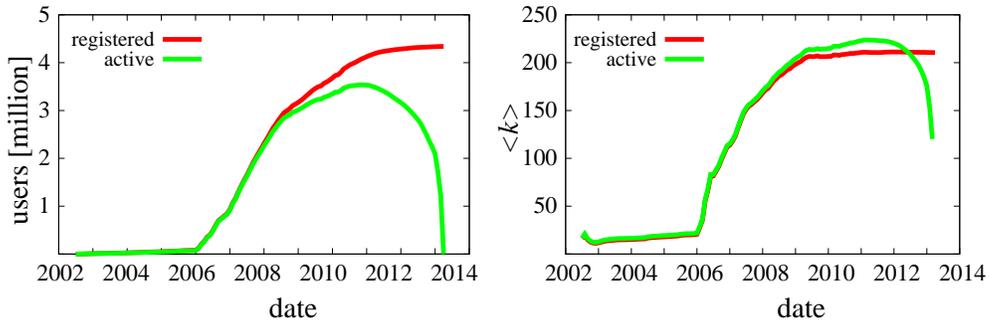}
\caption{\textbf{Timeline of properties of iWiW.} (a) The number of
registered users (red), the number of active users (green), (b)
average degree of the nodes, all registered (red), only among active
(green).
}
\label{Fig:avdeg} 
\end{figure}

Here we present results for iWiW \cite{iwiw}, the Hungarian OSN, which
was active between 2004 and 2013 and was in the list of the top three
most popular sites in the country for 2007-2011 \cite{webaudit}. The
details of the data are presented in the section Materials and
Methods. We carried out the same analysis for another social
network site: Gowalla, (see the section Other Empirical Networks in
Supporting Information).

\begin{figure}
\includegraphics[width=.95\textwidth]{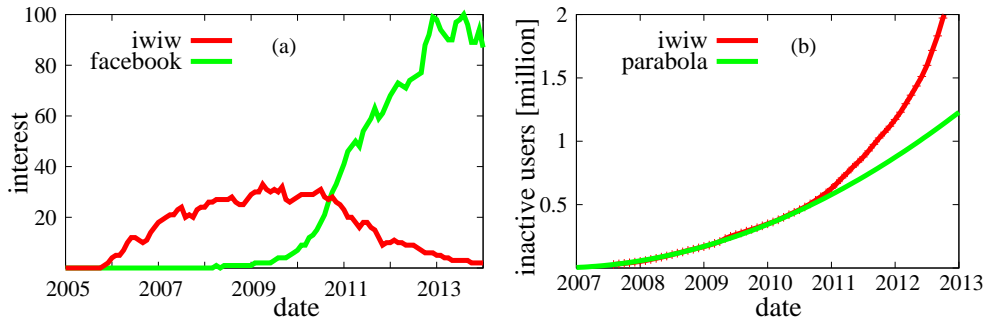}
\caption{\textbf{Time evolution of interest for social network sites in
Hungary and the inactive users on iWiW.} (a) Google trends of iWiW and
Facebook for Hungary.  (b) Number of users who became inactive in
until a given month. The green line is a parabola fit in the form of
$210m^2$, where $m$ is the number of months after July 2006.  Before
August 2010 this gives a perfect fit indicating a linearly increasing
churning rate. After that time the rate grows super linearly.}
\label{Fig:iWiWFacebook} 
\end{figure}

The  nodes of the network are the users of the service and links are
the mutually acknowledged connections ("friendships"). Figure
\ref{Fig:avdeg} shows the time evolution of the number of users of
iWiW as well as the average degrees for all and for the active nodes where
a node is considered to be inactive after its last login. After an
almost latent period the number of users started growing rapidly early
2006 (for reasons see Materials and Methods section) and until mid
2007 the number of leavers remained negligible.  It is interesting to
note that even though in Fig.~\ref{Fig:iWiWFacebook} (a) Google trends
shows that already in September 2010 the number of hits for Facebook
was higher than that for iWiW, the number of active
users of iWiW still increased for a few more months. The average
degree of all users remained constant after 2009 but the average
degree of the active users increased slightly till mid 2011 as shown
in Fig.~\ref{Fig:avdeg} (b), indicating that less embedded users left
first.

Figure~\ref{Fig:iWiWFacebook} (b) shows the cumulative number of users
who became inactive in a given month. Some users left the service
already as early as 2007 and from that point on the cumulative number
of inactive users increases quadratically, meaning that the number of
users leaving the service each month increases linearly with time.
This trend remained characteristic for the system for the period
2007-2010.  The observed linear increase in the rate of users leaving
the service is the result of an interplay between the slowly
growing popularity of Facebook and the increasing user pool.

\begin{figure}
\includegraphics[width=.95\textwidth]{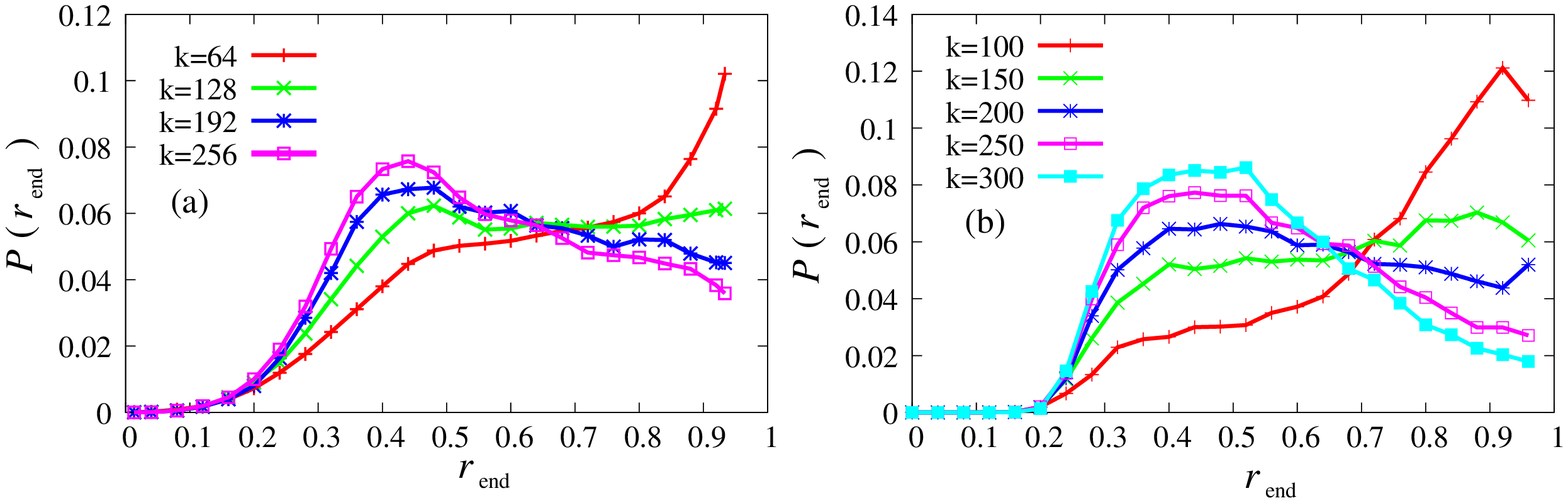}
\caption{\textbf{Fraction of active friends at the time of last
login.} (a) Distribution of the fraction of active acquaintances at
the time of the last login of users for four different user degrees in
iWiW. Note that the absence of values less than
$r_\mathrm{end}\simeq0.3$ is due to the fact that our database
was truncated when 70\% of the users left. (b) Same plot for the model
using an extended network with $N{=}10^4$ agents and $\langle k
\rangle {=}220$.
}
\label{Fig:fractionatleave} 
\end{figure}

Around the end of 2010 the number of users leaving iWiW started to
accelerate, which turned into a dramatic loss and lead finally
to the collapse of the entire OSN. The turning point in the history of
iWiW can be seen on the Google trends (Fig.~\ref{Fig:iWiWFacebook}
(a)) where the popularity of Facebook in Hungary started to increase
rapidly while the popularity of iWiW to decline, at the end of 2010.
The question is whether the dramatic increase in churning can be
simply explained by the increase in the interest of people in the
competing service or is there a collective effect. We will show that
the latter is the case.

In order to study the above question we calculated the fraction
$r_\mathrm{end}$ of the active acquaintances of a user at the time of
her last login with a week overhead. The results about the
distribution of $r_\mathrm{end}$ are shown in
Fig.~\ref{Fig:fractionatleave} (a) for users with different degree.

The distribution of the fraction of active friends at the time of last
login depends in a specific way on the number $k$ of friends of the
users. While the distribution is for all $k$-values smooth and spreads
over almost the whole range of $r_\mathrm{end}$, the position of the
maximum jumps from $\sim0.9$ for $k < 130$ to $0.4{-}0.5$ for $k \ge
130$ (see Supporting information for more details). This indicates the
following mechanism: For all $k$ values there are persons, who leave
the OSN just due to the exogenous input -- those, who have an active
friend ratio close to 1 are certainly such users as they have not been
influenced by the churning of their friends. Then, there are persons,
who leave, when a considerable part of their partners have left. We
can expect that the level of embeddedness into the OSN as measured by
the number of friends has an impact on which of these factors is
dominant. The users with the maximum close to 1 are mainly those with
low embeddedness in (or affinity to \cite{Torok_2016}) the service as
they have only relatively low number of friends in the OSN. On the
other hand users with high affinity have many friends on the OSN, and
they stick to the service as long as they can reach a good fraction of
their friends. The data suggests that for the latter group the maximum
of the distribution of active friends at the time of churning is
around 45\%. In the early period (mid 2007 -- end 2010) churning was
mainly due to independent decisions triggered by some exogenous input
or loss of interest. The deviation from the smooth, linearly
increasing rate indicates the appearance of the collective effects.

The mechanism of the sharp transition from one position of the maximum
to the other one can be understood such that the broad distributions
result from two main processes: One with a maximum close to 1 and the
other at around 0.5 and the weights of these get shifted from the
former to the latter one as the degree increases (see
Fig.~\ref{Fig:fractionatleave}).

\begin{figure}
\begin{center}
\includegraphics[width=.95\textwidth]{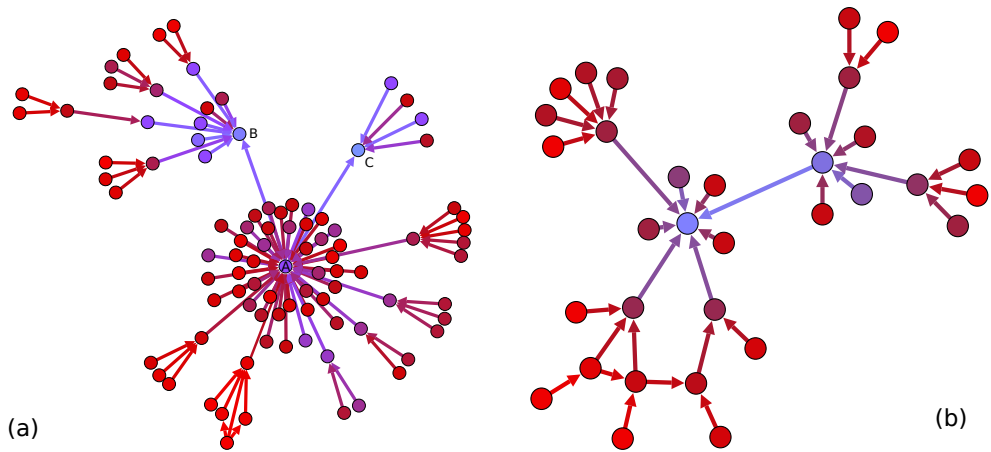}
\end{center}
\caption{\label{Fig:democascade} \textbf{Sample cascades.} Arrows mark links
between users who have acknowledged each other as friends and the user
at the tail of the arrow left not more than a week before the one at
the head. Colour changes from red to light blue as time passes.
(a) cascade based on iWiW data, note that after the leave of many
friends of A, it also left the site which then triggered the leave of
users B and C. The time evaluation of a cascade is shown in a movie in
the supplementary information; (b) model example cascade.
}
\end{figure}

The existence of thresholds in the decision to leave the service may
lead to collective phenomena of avalanches or cascades. Indeed, we
find such cascades in the data; for an example see
Fig.~\ref{Fig:democascade} (a).  A user was chosen who left the
service in March 2011. Arrows are plotted between friends if they left
the service in the given order (tail first) within one week.  Users
were traced back up to a two months period with the above condition
unfolding this way the process backwards.  There are  users, who seem
to leave the OSN without any endogenous trigger in this
representation.  On the other hand, there are clear examples of
cascades already on this one week scale, where users get inactive
after many of their friends got inactive. Thus the mechanism resulting
finally in a complete collapse of the OSN is as follows: At the
beginning people leave spontaneously the service due to lack of
interest or finding another provider more satisfactory. These are
usually individuals, who are not bound by many ties to the network. As
the interest in the competing service rises (cf.
Fig.~\ref{Fig:iWiWFacebook}~(a)) the number of such churners increases
such that the threshold level of persons with high degree is reached,
releasing further departures from the service and, occasionally,
cascades. However, the cascades are finite and need to be further
triggered by those who churn "spontaneously", i.e., due to exogenous
influence. This is similar to the situation recently observed for
innovation spreading \cite{Ruan_2015, Karsai_2016}, which gives
inspiration to modelling our empirical findings. 

\begin{figure}
\begin{center}
\includegraphics[width=.65\textwidth]{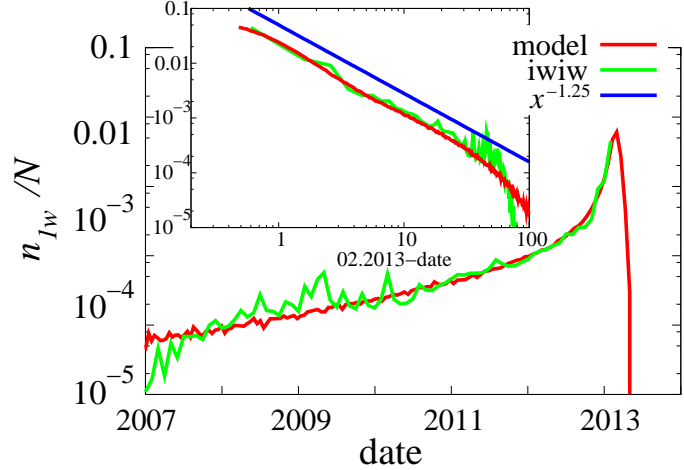}
\caption{\textbf{Time evolution of churning users.} The average number
$n_{1w}$ of friends churning during the week before the last login of
a user plotted against the time difference to the time of the maximal
user loss in case of the model and to the date February 2013 in case
of iWiW. The slope of the straight line is $\sim1.25$. The inset shows
the same data with the $x$ axis indicating the number of months after
01.2007.
}
\label{Fig:cascadefriend}
\end{center}
\end{figure}

We measured the average number of friends of a user churning
during the last 1,2, or 4 weeks before she left (denoted by
$n_{1w},~n_{2w},~n_{4w}$ respectively). Since apart from a constant
shift no important difference was found between these quantities only
$n_{1w}$ is plotted in Fig.~\ref{Fig:cascadefriend}, where we see the
following scenario: Users start to leave the OSN as early as in 2007
and has a strongly fluctuation period around $n_{1w} \approx 0.2$,
when the OSN is at its best. Beginning late 2010 $n_{1w}$ starts to
increase rapidly and finally it shoots up at early 2013, which can be
described as a power law with an exponent as $(t_c-t)^{-1.25}$ with $t_c=$~February 2013 (see inset of
Fig. \ref{Fig:cascadefriend}). We assume that this power law
divergence is a sign of the cascading collapse and can be used to
determine the date of the collapse beforehand.

A natural question to ask is whether there are other early signatures
of the global breakdown. For example, the statistics of newcomers and
of activity periods lead to interesting observations about the
healthiness of an the online service Myspace and other OSN-s
\cite{suvakov2012online,tadic2017agent}.
Here we analyse these two quantities as possible
indicators for the collapse of iWiW.

\begin{figure}
\begin{center}
\includegraphics[width=.95\textwidth]{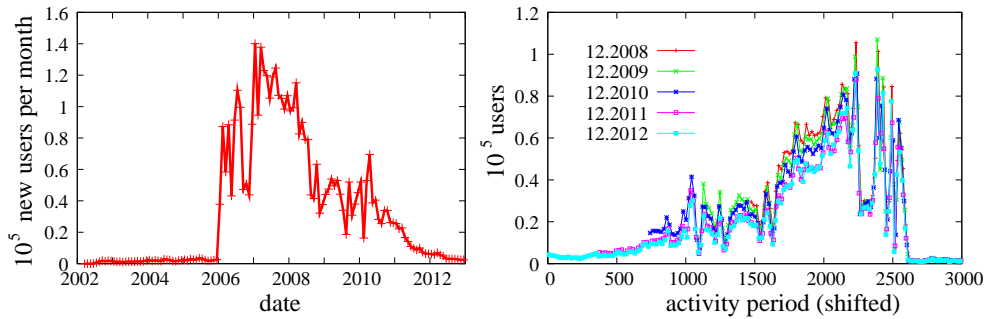}
\caption{\textbf{Time evolution of the new users and the activity
period of the active users:} (a) The number of new users per month
versus time. (b) The histogram of the activity period of the active
users. Each bin represents 20 days. The curves were shifted
successively by one year so that they would fall upon each other if
there was no change. The last curve 12.2012 was multiplied by a factor
of 2 for easier comparison with the previous ones.
}
\label{Fig:nudist} 
\end{center}
\end{figure}

In Fig.~\ref{Fig:nudist} (a) the number of new users per month is
plotted. One can easily recognise the commercialisation at the
beginning of 2006, the big capacity problem in September 2016. There
is a big drop in the number of new users in middle 2008. This is
related to the big crisis in Hungary. According to the data of
Worldbank \cite{Worldbank} the ratio of internet subscriptions
increased only 1\% in this year compared to the 8\% of the previous
one. From this period on the number of new users is less than half 
then previously. This period with 40-60 thousand new users
per month lasts till the end of 2011, when it slowly decreases to 0.
Obviously the big collapse of the site starts at around this point,
but this curve does not allow for an early recognition of this event.

We have seen in Fig.~\ref{Fig:fractionatleave} that those with low
degree leave the site first. In some online social sites the newcomers
replace the old ones (see e.g. gowalla) which keeps the site alive.
This hypothesis was tested on iWiW by measuring the distribution of
the activity period of the active users which is shown in
Fig.~\ref{Fig:nudist} (b). The curves prior to 2012 were shifted by
the number of days between the sample periods, so they fall on top of
each other if old users do not leave. Indeed we can see that
especially the part of the histograms for large activity period looks
very much time independent.  Even the big dip due to capacity problems
at the end of 2006 is the same on all curves. Note that the last curve
of 12.2012 was multiplied by a factor of 2 for better comparison,
because of the low number of active users by that time.  Still this
curve looks very much like the others.

The only difference we can see is that the last part of the older
curves are always higher which means that those who did not like the
site left not long after registering (most probably with low degree)
but the ones who stayed for a while remained almost to the end of the
site. The match by a factor of two between 12.2012 and 12.2011
indicates that the activity period was not a prime reason for staying
or leaving at the end, the collective motion indicated by the peak in
$r_{\mathrm{end}}$ seems more important.

\section*{Generalised threshold model and simulations}

In the following we construct a model for the collapse of OSN-s, which
incorporates both exogenous and collective effects on churning. The
parameters of the model will be adjusted to the iWiW data.

In order to keep the underlying structure simple, the model is defined
on a random network where 40\% of the links are created in an
uncorrelated random way and the rest is generated by placing random
triangles, as high clustering is typical for social
networks~\cite{kossinets2006empirical,WassermanFaust}. Finally we have
a random network with $N$ nodes and an average degree of $\langle
k\rangle$. At the beginning all nodes are part of the OSN. In the
Supplementary Information we present results using part of iWiW
network as underlying structure.

Social contacts have different intimacy levels
building a hierarchy of concentric layers around the persons in their
egocentric networks~\cite{Dunbar_layers}. It is natural to assume that
only the set of closest friends matter in the decision of well
embedded individuals about staying with the service or leaving it.
Therefore we will consider only a network with reduced average degree,
which is considered as a parameter. 

For the endogenous, collective effects we implement a threshold
mechanism \cite{Granovetter_1978,Watts_2002}: Whenever the fraction of
active friends of a node drops below its threshold $\lambda$ the node
will feel inclined to leave the OSN. This does not happen immediately
but with a rate $1/\tau$, so $\tau$ is the timescale of leaving the
OSN after the threshold condition is fulfilled.  All nodes are given a
predefined uncorrelated threshold value $\lambda=0.5\pm0.2$ with
uniform variations. 
 
The spontaneous churning is defined as follows: At each timestep nodes
leave the service with a rate linearly increasing with time:
$\gamma=\mu t/\tau$, reflecting the growing interest in the competing
service. We select nodes to leave the service with the probability
which was a decreasing function of the node degree (see Supporting
Information for details).  This was motivated by
the fact that more embedded users are more reluctant to leave the site
due to exogenous effects.

In summary the model is defined as follows:
\begin{enumerate}
\item A network with average degree $\langle k \rangle$ representing
the friendship is created with all users begin member of the OSN.
\item At time $t$ users leave with rate $\gamma=\mu t/\tau$. (For the
reproduction of $k$ dependence of $r_\mathrm{end}$ a $k$ dependence in
node selection may be introduced)
\item Users for which the ratio of the active friends dropped below
the threshold $\lambda$ are moved to the \textit{leaving} queue
\item Users in the leaving queue leave definitely with rate $\tau$. If
this happens check all its friends for threshold
\end{enumerate}

The dynamics defined this way has two timescales: $\tau$ and $1/\mu$.
The latter determines the exogenous timescale and can be fitted to
the initial quadratic increase of the number of inactive agents, which
relates the model time to real time, therefore it has no influence on the
dynamics. The value of the threshold can be adjusted by measuring the
peak in the $r_\mathrm{end}$ curves.

\begin{figure}
\begin{center}
\includegraphics[width=.95\textwidth]{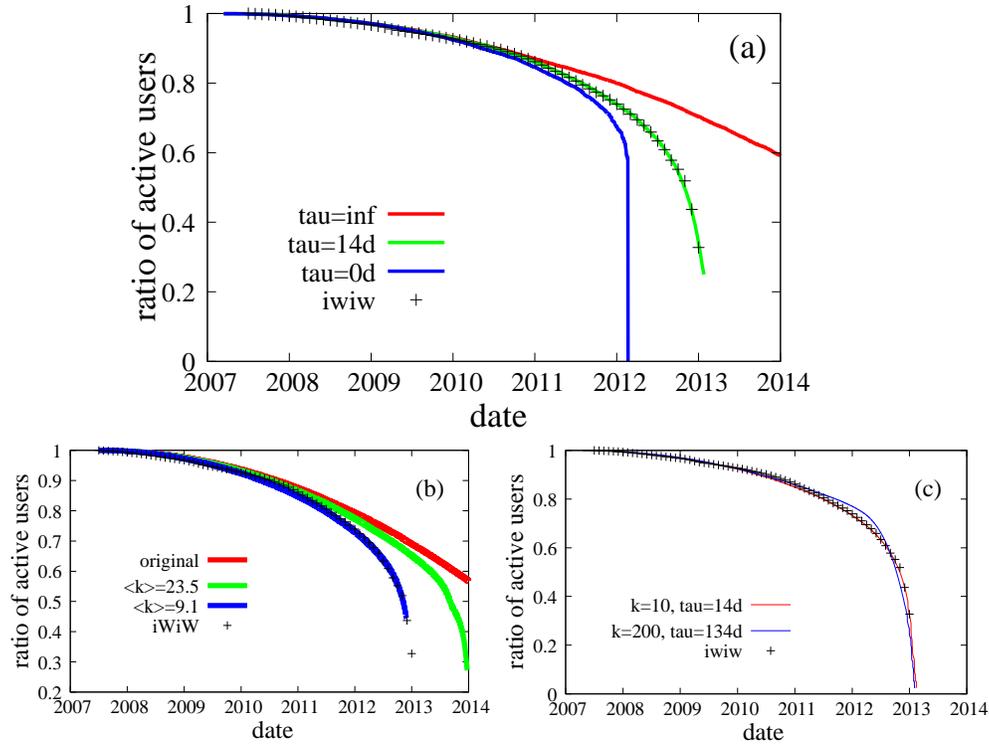}
\caption{\textbf{Time evolution of the active users in iWiW.} (a) The
cumulative fraction of active users (all users having a last login
date after the indicated time) in iWiW (crosses) and in the cascade
model (green). For comparison we show also results with zero
characteristic leaving time (blue), and with zero threshold, i.e. no
social effect (red).  Parameters: $N{=}10000$, $\langle k
\rangle{=}10$.
(b) The cumulative fraction of active users in iWiW and in the model
using a county as the underlying network of the model. (c) Model fits
using random network with average degree $\langle k \rangle=10$,
$200$.
}
\label{Fig:modelfit} 
\end{center}
\end{figure}

We are left with two parameters: $\tau$ and $\langle k\rangle$.
The timescale $\tau$ controls the speed of the collapse of the
service together with the average degree of the original network. If
either $\tau$ is small (it was zero in the original threshold model
\cite{Watts_2002}) of the network has a high average degree the
collapse of the site is almost instantaneous. We found that either
large waiting time or low degree is needed to recover the empirical
results.

Fitting our model to the empirical data gives a prefect match as shown
in Fig.~\ref{Fig:modelfit} (a) with values of $\langle k\rangle{=}10$
and $\tau=14.5$ days. Both values are realistic: First,  it is natural
to assume that people check back for two more weeks after they start
getting motivated to leave the service.  Second, it was already
suggested in \cite{Dunbar_layers} that people have a circle of
intimacy of about $12{-}15$ friends and relatives with whom they have
regular communication; these are the relationships, which are
particularly important for them. In Fig.~\ref{Fig:modelfit} (b) we
show the best fit with $\langle k\rangle{=}200$. It is clear visually
that this is a worse fit. For more details see Supporting Information.

Here we used an artificial random network for the model
studies. Since the contact network is available from iWiW we may use
part of it (the whole being too large for simulation) to study the
model. It was shown that political borders inside Hungary are apparent
through the contact network \cite{Jakobi_2014} therefore we have
decided to use the the users in one of the counties to simulate
the model.

We have introduced a probability with which the links are kept in
order to be able to decrease the degree of the network. After
decimating the links we remove nodes with degree 0 and 1.  We have
studied the evolution of the active users and the results is shown in
Fig.~\ref{Fig:modelfit} (c).  The best fit was obtained with an
average degree 9.1 similar to the model network.

In Fig.~\ref{Fig:modelfit} we have also shown the time evolution of
the ratio of the active users without avalanches and with no waiting
time. It can be seen that the time evolution of the three curves
depart at middle 2010 where the concurrent service becomes popular and
collective effects become important. The model generates cascades similar to those observed in the empirical data, see Fig.~\ref{Fig:democascade} (b).

A further verification of the model besides Fig.~\ref{Fig:democascade}
can be obtained by calculating the degree dependence of the number of
churned neighbours at leaving the network
Fig.~\ref{Fig:fractionatleave} (a). In order to compare the results of
the model with the empirical ones we scaled up the model in the
following way: We introduced further (socially less important) links
such that the average degree became the same as in the OSN (220). The
links were attached to the nodes with a probability proportional to
the degree in the small network and the triadic closures were also
applied with the same density. The similarity between
Fig.~\ref{Fig:fractionatleave} (a) and (b) are rather convincing. 

We have also calculated the average number of churning users as function of
the elapsed time. The results match well with the empirical data
as shown in Fig.~\ref{Fig:cascadefriend}.

\section*{Summary}

We studied the collapse of an online social network both empirically
and by modelling. The empirical results suggest that in the early
stage users leave the OSN mostly because they are not anchored to it
by a large number of friends, that means, they are not
interested enough in online networking or they churn to a competing
service.  As time goes on, the latter effect seems to dominate
resulting in a linearly increasing churning rate following the
increasing popularity of the competitor. This erosion of the network
due to exogenous influence has the effect that the social pressure 
becomes critical for an increasing number of active users, who then
decide to leave, which leads to a cascading effect and finally, to the
collapse of the OSN.

The endogenous and exogenous effects were incorporated in a threshold
model, which could reasonably well reproduce the measured quantities
of the empirical system. Starting from a recent
generalisation~\cite{Ruan_2015,Karsai_2016} of the Watts threshold
model~\cite{Watts_2002} the most important new elements of our model
are as follows: First, we have taken into account the endogenous
influence by introducing spontaneous churning with a time dependent
rate, reflecting the increasing interest in the competitor OSN.
Second, we introduced a delay time $\tau$ between the fulfilment of
the threshold criterion and the action under the social pressure. This
realistic element was used as a fitting parameter in the model
calculations.

The spontaneous churning
due to endogenous influence dominates the early phase of the OSN,
however, gradually cascades occur and finally they become dominant.
From that time any effort to damp spontaneous churning (e.g., by
advertisement) is in vain. The apparent stability of the OSN was only
due to the relatively large characteristic time for cascading users.

We have found empirically and reproduced by the model that the number
of churning neighbours of a churning user within a given time before
the event behaves as a power law with a critical time (February 2013,
in our case), which can be considered as the collapse time of the
service. Similar observations in other systems may be used to predict
the date of collapse due to social pressure.

\section*{Methods}

\subsection*{Data.}
\label{Sec:data}

The social network iWiW (4.3 million registered users) was started in
2002 and closed down June 2014. The data used for this study for the
period of April 4, 2002 -- January 23, 2013 includes: date of
registration, last login of each user and time stamped link creation
information. Using this data we can reconstruct the whole life cycle
of iWiW (see Fig~\ref{Fig:avdeg}). It was at its peak the second most
popular site in Hungary \cite{webaudit} it had 3.5 million active
users in 2010, in a country with a population of 10 million (worldwide
13 million native speakers) and at that time about 60\% Internet
penetration \cite{Eurostat,Worldbank}. The site started as a
non-profit project in April 2002, however, it was purchased by a giant
telecom company in 2006 and remained the leading social network site
of Hungary for years; in fact, it was considered as a main driving
force behind the speed up of Internet penetration in Hungary. 

Till middle 2011 iWiW was invitation based. Every user after 30-50
days of waiting time has got one voucher and new users could register
only if they received a voucher from an already existing member. Later
vouchers were redistributed irregularly until, in the last period
after 2012, the registration became unconditional. 

The site has remained widely used even after Facebook became popular
worldwide \cite{Jakobi_2014}, (see Fig.~\ref{Fig:iWiWFacebook}). The
story of iWiW came to a sudden end due to various reasons. i) the
introduction of games and application in Facebook made it more
attractive especially to young people, ii) the lack of usable message
filtering system made iWiW a prime target of spammers using mainly
compromised accounts, iii) a sizeable ($\sim$100 thousand) Hungarians
living abroad also gave a strong push to convert friends to Facebook,
which rapidly became the prime Hungarian social network site. This
resulted in a rapid increase in the number of churning users in 2011
and lead finally to a collapse in 2012.  The site was closed down in
June 2014 after passing "time capsules" with their network data to the
registered users.

The data we used contains anonymised users with registration and last
login date, and a time stamped connection list. We used the following
data items: timestamp of the last login of a user. The list of friends
for each user from the connections which were mutually acknowledged.

In addition to iWiW, our main focus here, we have also studied a
much smaller network, available from the Large Stanford Dataset
Collection \cite{snapnets} and which is location based social network
site. Gowalla, established in 2007 had $\sim$600.000
users in 2010; it was acquired by Facebook in 2011 and closed down
March 2012. The necessary dynamic information to reconstruct the life
cycle of the network was available. The results of this network is
presented in the Supporting Information. 

\subsection*{Data evaluation.}
\label{Sec:evaluation}

Using the last login timestamp and friendship network
Figs.~\ref{Fig:avdeg} and \ref{Fig:iWiWFacebook} (b) were trivially
obtained. When counting the number of active friends of a user at the
time of its last login we did not consider ordering dates less than 15
days apart. The reason behind this decision was the weekly usage of
the site and the similar characteristic time of the churning. In the
evaluation of $r_\mathrm{end}$ of the model results we used the same
code as for the iWiW data.

Cascade users were defined as number of acquaintances having a last
login less than one (two, or four) week(s) before the last login date
of the user.

Supplementary movie was made as follows: Movie time goes proportional
to real time, Yellow circles indicate the last login of a given user.
An arrow is drawn if the head user leaves no later than a week after
the last login of the user at the tail. Colour goes from red to light
blur with elapsed time.


\section*{Acknowledgement}

This work was supported by OTKA K112713 of the Hungarian
Science foundation and by the H2020 FETPROACTGSS CIMPLEX Grant No.
641191. We thank Zhongyuan Ruan for data analysis at an early stage
of this work.

\section*{Author contributions}

Authors (J.T. and J.K.) declare equal 50-50\% contribution to the article.

\section*{Competing financial interests}

The authors declare no competing financial interests.

\bibliographystyle{plain}
\bibliography{l2}

\begin{thebibliography}{10}

\bibitem{Albert_2000}
R{\'e}ka Albert, Hawoong Jeong, and Albert-L{\'a}szl{\'o} Barab{\'a}si.
\newblock Error and attack tolerance of complex networks.
\newblock {\em nature}, 406(6794):378--382, 2000.

\bibitem{andjelkovic2015hierarchical}
Miroslav Andjelkovi{\'c}, Bosiljka Tadi{\'c}, Slobodan Maleti{\'c}, and Milan
  Rajkovi{\'c}.
\newblock Hierarchical sequencing of online social graphs.
\newblock {\em Physica A: Statistical Mechanics and its Applications},
  436:582--595, 2015.

\bibitem{Barabasi_1999}
Albert-L{\'a}szl{\'o} Barab{\'a}si and R{\'e}ka Albert.
\newblock Emergence of scaling in random networks.
\newblock {\em science}, 286(5439):509--512, 1999.

\bibitem{Buldyrev_2010}
Sergey~V Buldyrev, Roni Parshani, Gerald Paul, H~Eugene Stanley, and Shlomo
  Havlin.
\newblock Catastrophic cascade of failures in interdependent networks.
\newblock {\em Nature}, 464(7291):1025--1028, 2010.

\bibitem{Centola_2010}
Damon Centola.
\newblock The spread of behavior in an online social network experiment.
\newblock {\em Science}, 329(5996):1194--1197, 2010.

\bibitem{Cohen_2000}
Reuven Cohen, Keren Erez, Daniel Ben-Avraham, and Shlomo Havlin.
\newblock Resilience of the internet to random breakdowns.
\newblock {\em Physical review letters}, 85(21):4626, 2000.

\bibitem{Cohen_2001}
Reuven Cohen, Keren Erez, Daniel Ben-Avraham, and Shlomo Havlin.
\newblock Breakdown of the internet under intentional attack.
\newblock {\em Physical review letters}, 86(16):3682, 2001.

\bibitem{Dunbar_layers}
Robin~IM Dunbar, Valerio Arnaboldi, Marco Conti, and Andrea Passarella.
\newblock The structure of online social networks mirrors those in the offline
  world.
\newblock {\em Social Networks}, 43:39--47, 2015.

\bibitem{Easley_2010}
David Easley and Jon Kleinberg.
\newblock {\em Networks, crowds, and markets: Reasoning about a highly
  connected world}.
\newblock Cambridge University Press, 2010.

\bibitem{Eurostat}
Eurostat.
\newblock Households having access to the internet by type of connection,
  2007-2013.

\bibitem{Gleeson_2007}
James~P Gleeson and Diarmuid~J Cahalane.
\newblock Seed size strongly affects cascades on random networks.
\newblock {\em Physical Review E}, 75(5):056103, 2007.

\bibitem{Gleeson_2011}
J.P. Gleeson.
\newblock High-accuracy approximation of binary-state dynamics on networks.
\newblock {\em Physical Review Letters}, 107:068701, 2011.

\bibitem{Gleeson_2013}
J.P. Gleeson.
\newblock Binary-state dynamics on complex networks: Pair approximation and
  beyond.
\newblock {\em Physical Review X}, 3:021004, 2013.

\bibitem{Goltsev_2006}
Alexander~V Goltsev, Sergey~N Dorogovtsev, and JFF Mendes.
\newblock k-core (bootstrap) percolation on complex networks: Critical
  phenomena and nonlocal effects.
\newblock {\em Physical Review E}, 73(5):056101, 2006.

\bibitem{Gonzales-Baylon_2011}
Sandra Gonz{\'a}lez{-}Bail{\'o}n, Javier Borge{-}Holthoefer, Alejandro Rivero,
  and Yamir Moreno.
\newblock The dynamics of protest recruitment through an online network.
\newblock {\em Scientific reports}, 1, 2011.

\bibitem{Granovetter_1978}
Mark Granovetter.
\newblock Threshold models of collective behavior.
\newblock {\em American journal of sociology}, pages 1420--1443, 1978.

\bibitem{Karsai_2016}
M.~Karsai, Gerardo I{\~n}iguez, Riivo Kikas, K.~Kaski, and J.~Kert\'esz.
\newblock Local cascades induced global contagion: How heterogeneous
  thresholds, exogenous effects, and unconcerned behaviour govern online
  adoption spreading.
\newblock {\em Scientific Reports}, 6:27178, 2016.

\bibitem{Karsai_2014}
M{\'a}rton Karsai, Gerardo I{\~n}iguez, Kimmo Kaski, and J{\'a}nos Kert{\'e}sz.
\newblock Complex contagion process in spreading of online innovation.
\newblock {\em Journal of The Royal Society Interface}, 11(101):20140694, 2014.

\bibitem{webaudit}
Gemius~Hungary Kft. and Ipsos.
\newblock gemius/ipsos audience.
\newblock \url{http://www.audience.gemius.hu/}, 2007-2013.

\bibitem{Kim_2004}
Hee-Su Kim and Choong-Han Yoon.
\newblock Determinants of subscriber churn and customer loyalty in the korean
  mobile telephony market.
\newblock {\em Telecommunications Policy}, 28:751–765, 2004.

\bibitem{Kleinberg_1999}
Jon~M Kleinberg.
\newblock Authoritative sources in a hyperlinked environment.
\newblock {\em Journal of the ACM (JACM)}, 46(5):604--632, 1999.

\bibitem{kossinets2006empirical}
Gueorgi Kossinets and Duncan~J Watts.
\newblock Empirical analysis of an evolving social network.
\newblock {\em science}, 311(5757):88--90, 2006.

\bibitem{Jakobi_2014}
Bal{\'a}zs Lengyel, Attila Varga, Bence S{\'a}gv{\'a}ri, {\'A}kos Jakobi, and
  J{\'a}nos Kert{\'e}sz.
\newblock Geographies of an online social network: weak distance decay effect
  and strong spatial modularity.
\newblock {\em PLOS ONE}, 2014.

\bibitem{Lerman_2010}
Kristina Lerman and Rumi Ghosh.
\newblock Information contagion: An empirical study of the spread of news on
  digg and twitter social networks.
\newblock {\em ICWSM}, 10:90--97, 2010.

\bibitem{Leskovec_2005}
Jure Leskovec, Jon Kleinberg, and Christos Faloutsos.
\newblock Graphs over time: densification laws, shrinking diameters and
  possible explanations.
\newblock In {\em Proceedings of the eleventh ACM SIGKDD international
  conference on Knowledge discovery in data mining}, pages 177--187. ACM, 2005.

\bibitem{snapnets}
Jure Leskovec and Andrej Krevl.
\newblock {SNAP Datasets}: {Stanford} large network dataset collection.
\newblock \url{http://snap.stanford.edu/data}, June 2014.

\bibitem{mitrovic2011quantitative}
Marija Mitrovi{\'c}, Georgios Paltoglou, and Bosiljka Tadi{\'c}.
\newblock Quantitative analysis of bloggers’ collective behavior powered by
  emotions.
\newblock {\em Journal of Statistical Mechanics: Theory and Experiment},
  2011(02):P02005, 2011.

\bibitem{Morone_2015}
Flaviano Morone and Hern{\'a}n~A Makse.
\newblock Influence maximization in complex networks through optimal
  percolation.
\newblock {\em Nature}, 2015.

\bibitem{Newman_2010}
Mark Newman.
\newblock {\em Networks: an introduction}.
\newblock Oxford university press, 2010.

\bibitem{Pastor-Satorras_2015}
Romualdo Pastor{-}Satorras, Claudio Castellano, Piet Van~Mieghem, and
  Alessandro Vespignani.
\newblock Epidemic processes in complex networks.
\newblock {\em Reviews of modern physics}, 87(3):925, 2015.

\bibitem{Rogers_2003}
Everett~M Rogers.
\newblock Elements of diffusion.
\newblock {\em Diffusion of innovations}, 5:1--38, 2003.

\bibitem{Ruan_2015}
Zhongyuan Ruan, Gerardo Iniguez, M{\'a}rton Karsai, and J{\'a}nos Kert{\'e}sz.
\newblock Kinetics of social contagion.
\newblock {\em Physical review letters}, 115(21):218702, 2015.

\bibitem{WassermanFaust}
Wasserman S. and K~Faust.
\newblock {\em Social Network Analysis: Methods and Applications.}
\newblock Cambridge University Press, Cambridge, UK, 1994.

\bibitem{Saavedra_2008}
Serguei Saavedra, Felix Reed-Tsochas, and Brian Uzzi.
\newblock Asymmetric disassembly and robustness in declining networks.
\newblock {\em Proceedings of the National Academy of Sciences},
  105(43):16466--16471, 2008.

\bibitem{Singh_2013}
Pramesh Singh, Sameet Sreenivasan, Boleslaw~K Szymanski, and Gyorgy Korniss.
\newblock Threshold-limited spreading in social networks with multiple
  initiators.
\newblock {\em Scientific reports}, 3, 2013.

\bibitem{suvakov2012online}
Milovan Suvakov, Marija Mitrovic, Vladimir Gligorijevic, and Bosiljka Tadic.
\newblock How the online social networks are used: Dialogs-based structure of
  myspace.
\newblock {\em arXiv preprint arXiv:1206.6588}, 2012.

\bibitem{tadic2013co}
Bosiljka Tadi{\'c}, Vladimir Gligorijevi{\'c}, Marija Mitrovi{\'c}, and Milovan
  {\v{S}}uvakov.
\newblock Co-evolutionary mechanisms of emotional bursts in online social
  dynamics and networks.
\newblock {\em Entropy}, 15(12):5084--5120, 2013.

\bibitem{tadic2017agent}
Bosiljka Tadi{\'c}, Milovan {\v{S}}uvakov, David Garcia, and Frank Schweitzer.
\newblock Agent-based simulations of emotional dialogs in the online social
  network myspace.
\newblock In {\em Cyberemotions}, pages 207--229. Springer, 2017.

\bibitem{Worldbank}
{\relax The World Bank}.
\newblock Internet users, 1990-2014.

\bibitem{Torok_2016}
J{\'a}nos T{\"o}r{\"o}k, Yohsuke Murase, Hang-Hyun Jo, J{\'a}nos Kert{\'e}sz,
  and Kimmo Kaski.
\newblock What does big data tell? sampling the social network by communication
  channels.
\newblock {\em Physical Review E}, 94:052319, 2016.

\bibitem{Watts_2002}
Duncan~J Watts.
\newblock A simple model of global cascades on random networks.
\newblock {\em Proceedings of the National Academy of Sciences},
  99(9):5766--5771, 2002.

\bibitem{Weng_2013}
Lilian Weng, Jacob Ratkiewicz, Nicola Perra, Bruno Gon{\c{c}}alves, Carlos
  Castillo, Francesco Bonchi, Rossano Schifanella, Filippo Menczer, and
  Alessandro Flammini.
\newblock The role of information diffusion in the evolution of social
  networks.
\newblock In {\em Proceedings of the 19th ACM SIGKDD international conference
  on Knowledge discovery and data mining}, pages 356--364. ACM, 2013.

\bibitem{iwiw}
Wikipedia.
\newblock iwiw --- {W}ikipedia{,} the free encyclopedia, 2016.
\newblock [Online; accessed 10-Aug-2016].

\end{thebibliography}

\end{document}